# Electronic and Superconducting Properties of Some FeSe-based Single Crystals and Films Grown Hydrothermally

**Xiaoli Dong[1, 2, 3*], Fang Zhou[1, 2, 3], and Zhongxian Zhao[1, 2, 3]**

[1] Beijing National Laboratory for Condensed Matter Physics and Institute of Physics, Chinese Academy of Sciences, Beijing 100190, China

[2] University of Chinese Academy of Sciences, Beijing 100049, China

[3] Songshan Lake Materials Laboratory, Dongguan, Guangdong 523808, China

**\* Correspondence to:** dong@iphy.ac.cn

**Keywords: iron selenide superconductors, superconducting and normal-state properties, spin nematicity, electronic phase separation, high critical parameters, and hydrothermal growth.**

## Abstract

Our recent year's studies of the prototypal FeSe and molecule-intercalated (Li,Fe)OHFeSe superconductor systems are briefly reviewed here, with emphasis on experimental observations of the link between the superconductivity and normal-state electronic property in the single crystals and films. These samples were successfully synthesized by our recently developed soft-chemical hydrothermal methods, which are also briefly described. Particularly in the Mn-doped high-$T_c$ (Li,Fe)OHFeSe film, a strong enhancement of the superconducting critical current density was achieved, which is promising for practical application of the superconductivity.

## 1. Introduction

Iron-based superconductors [1] have received extensive attention because of their rich physics, including magnetic and nematic instabilities, electronic correlations, and quantum phenomena [2-9]. As the second class of high-$T_C$ materials after the discovery of cuprate superconductors, the iron-based superconductors are also promising for practical application owing to their large critical current density, high upper critical field, and small anisotropy [10-17]. The recent observation of Majorana zero modes in iron-based superconductors implies a potentiality for future application in topological quantum calculating [18-21]. Unlike an electronic configuration of Cu-3d$^9$ in the cuprates, the iron-based compounds have an electronic configuration of Fe-3d$^6$ and a small crystal-field splitting [2,7,22-24]. An immediate consequence of this is that all the five Fe-3d orbitals could be involved in the low-energy interactions [25], giving rise to the multiband nature of the iron-based superconductivity, and the complexity and multiplicity of the normal-state properties. The iron-based family has two major subclasses, the iron chalcogenide and pnictide superconductors. Among them, the iron selenide superconductors have been shown to display a highly tunable superconducting critical $T_c$ and unique electronic properties in the normal state, thus providing a superior platform to investigate the underlying physics for iron-based superconductivity.

Superconductivity of FeSe-based compounds emerges from the layered blocks formed by edge-sharing FeSe-tetrahedra, each containing one iron-plane sandwiched between two selenium-planes. An important feature is that the superconducting $T_c$ can be tuned in a wide range. The simplest binary FeSe shows bulk superconductivity at a lower $T_c \sim 9$ K under ambient pressure [26]. It is notable that $T_c$ can be boosted to tens of kelvin (30 K $-$ 50K), by the applications of high pressure [27-33], charge-carrier injection[34], electrochemical etching [35], and chemical intercalation. The weak van der Waals bonding between the neighboring FeSe-blocks allows a variety of FeSe-based intercalates to be obtained, such as the atom-intercalated $A_y Fe_{2-x} Se_2$ (A=alkali metal) [36-40], molecule-intercalated (Li$_{0.8}$Fe$_{0.2}$)OHFeSe [41], and atom/molecule-co-intercalated Li$_x$(C$_5$H$_5$N)$_y$Fe$_{2-z}$Se$_2$ [42], A$_x$(NH$_2$)$_y$(NH$_3$)$_{1-y}$Fe$_2$Se$_2$ [43,44], A$_x$(NH$_3$)$_y$Fe$_2$Se$_2$ [45-47] and A$_x$(C$_2$H$_8$N$_2$)$_y$Fe$_2$Se$_2$ [48]. Moreover, the highest superconducting gap opening temperature (~65 K) among all the iron-based superconductors has been observed in a monolayer FeSe grown on a SrTiO$_3$ substrate [49,50]. On the other hand, distinct from most iron-based superconductor systems, FeSe does not order magnetically at ambient pressure. However, a unique electronic nematic order has been observed to develop in FeSe, which is associated with a rotational-symmetry-breaking transition from a tetragonal to an orthorhombic phase at $T_s \sim 90$ K [51,52]. The electronic nematicity is directly related to a degeneracy lifting of the bands with Fe 3d$_{xz}$ and 3d$_{yz}$ orbital characters [53-55]. Compared to the Fermi-surface topology of the prototypal FeSe, in the molecule-intercalated (Li,Fe)OHFeSe single crystals, only the electron pockets near the Brillouin zone corners are observed, in absence of the hole pocket near the zone center [56,57]. This raises question about a proposed pairing scenario of the electronic scatterings between the hole-like and electron-like pockets. Study of the FeSe-based superconductors is essential for a better understanding of the unconventional superconductivity.

To investigate the link between the unconventional superconductivity and unusual normal-state electronic properties, and the potential for the superconductivity application, high-quality single crystal and film samples are highly demanded. Recent years, we have been exploring soft-chemical methods suitable for synthesizing the FeSe-based superconductor single crystals and single-crystalline films hard to obtain by conventional high-temperature growth. By developing hydrothermal ion-exchange [58-60] and ion-deintercalation [61,62] approaches, we have succeeded in synthesizing series of high-quality sizable single crystals of the intercalated (Li,Fe)OHFeSe and binary FeSe systems, respectively. Our further study [9] has shown a strong electronic two-dimensionality and a nearly linear extracted magnetic susceptibility in the hydrothermal high-$T_c$ (42 K) (Li,Fe)OHFeSe single crystal, suggesting the presence of two-dimensional magnetic fluctuations in the normal state. In a series of the (Li, Fe)OHFeSe single crystals, a coexistence of antiferromagnetism with superconductivity has been detected [60]. We explain such coexistence by electronic phase separation, similar to the previously observed in high-$T_c$ cuprates and iron arsenides. An electronic phase diagram is further established for (Li, Fe)OHFeSe system [60,63]. In hydrothermal binary $Fe_{1-x}Se$ single crystals, we have observed a field-induced two-fold rotational symmetry emerging below $T_{sn}$ in angular-dependent magnetoresistance measurements, and a linear relationship between $T_c$ and $T_{sn}$ [61,64]. Importantly, we find in our recent study [9] that the superconductivity of FeSe system emerges from the strongly correlated, hole-dominated $Fe_{1-x}Se$ as the non-stoichiometry is reduced to $x \sim 5.3$ %. Interestingly, such an $x$ threshold for superconductivity of the prototypal FeSe is similar to that ($x \sim 5$ % [65]) for high-$T_c$ superconductivity of the intercalated (Li, Fe)OHFeSe sharing the common superconducting FeSe-blocks.

We have also succeeded in synthesizing a series of high-quality single-crystalline films of (Li, Fe)OHFeSe system, by inventing a hydrothermal epitaxial film technique [16,17,66]. We find that doping Mn into high-$T_c$ (Li, Fe)OHFeSe films can raise the superconducting critical current density $J_c$ by one order of magnitude to 0.32 MA/cm$^2$ at a high field of 33 T [17]. Such a high $J_c$ value is the record so far among the iron-based superconductors, and is thus promising for high-field application of the superconductivity. Besides, our breakthrough in the crystal growth has greatly promoted other related studies and progresses have been made [57,59,60,67-71], including the ARPES study of Fermi-surface topology [57] and the observation of pressure-induced second high-$T_c$ (> 50 K) phase [70] in the (Li,Fe)OHFeSe system. Our developed growth method has also been adopted in the studies of other research groups [56,72-83].

## 2. Soft-chemical hydrothermal growth methods developed for FeSe-based single crystals and films

The discovery of $Li_{0.8}Fe_{0.2}OHFeSe$ (FeSe-11111) superconductor [41] brings new opportunity for the study of iron-based superconductivity. (Li, Fe)OHFeSe is free from the complications of the structural transition, associated with the electronic nematicity, and the chemical phase separation, related to the inter-grown insulating $K_{0.8}Fe_{1.6}Se_2$ (KFS-245 phase) [63], as compared to the prototypal FeSe-11 and $K_{1-y}Fe_{2-x}Se_2$-122 superconductors, respectively. Moreover, it shows an ambient-pressure high $T_c$ = 42 K and a pressure-induced higher $T_c$ > 50 K under 12.5 GPa [70]. Having a Fermi-surface topology [56,57] similar to the high-$T_c$ (>65 K) FeSe monolayer, (Li, Fe)OHFeSe system turns out to be an ideal platform for studying the superconducting and normal-state properties of high-$T_c$ iron-based superconductors. Initially, only the powder samples of (Li, Fe)OHFeSe can be prepared hydrothermally [41,63,65,84,85]. For in-depth investigations on the intrinsic and anisotropic physical properties, the high-quality single crystal and film samples are indispensable.

The crystal structure of (Li, Fe)OHFeSe consists of a stacking of one superconducting (SC) FeSe-block alternating with one insulating (Li, Fe)OH-block along the $c$-axis. The (Li, Fe)OHFeSe compound suffers an easy decomposition because of the inherent weak hydrogen bonding. Therefore, none of the conventional high-temperature methods is applicable to grow the single crystals. To overcome this problem, we have developed a soft-chemical hydrothermal ion-exchange method capable of producing high-quality sizable single crystals of (Li, Fe)OHFeSe [58]. For the hydrothermal ion-exchange reaction, large and high-quality $K_{0.8}Fe_{1.6}Se_2$ crystal is used as a kind of matrix. The structure of $K_{0.8}Fe_{1.6}Se_2$ is formed by alternatively stacked K-layer and FeSe-tetrahedron-block similar to the target compound. The K ions in $K_{0.8}Fe_{1.6}Se_2$ are completely de-intercalated during the hydrothermal process. Simultaneously, the (Li, Fe)OH-blocks constructed by ions from the hydrothermal solution are intercalated into the matrix, and the ordered vacant Fe-sites (20% in amount) originally in the matrix $Fe_{0.8}Se$-blocks are almost occupied. A series of large and high-quality (Li, Fe)OHFeSe single crystals [60] are thus derived. Fig. 1 schematically illustrates the hydrothermal ion-exchange process. The derived (Li, Fe)OHFeSe single crystal almost inherits the original shape of the matrix (insets of Fig. 1b, c). Inspired by the successful hydrothermal ion-exchange method for the single crystals, we have further invented a hydrothermal epitaxial film technique to fabricate a series of high-quality single-crystalline films of un-doped [66] and Mn-doped [17] (Li, Fe)OHFeSe showing an optimal zero-resistivity $T_c$ = 42.4 K. This series of high-quality (Li, Fe)OHFeSe films has enabled a systematic study of the superconducting and normal-state properties [66].

By modifying the hydrothermal reaction conditions, we have also developed a hydrothermal ion-deintercalation (HID) method, as illustrated in Fig. 2. The atomic ratio of the FeSe-blocks can be continuously tuned by the HID process, yielding a series of non-stoichiometric $Fe_{1-x}Se$ single crystals at various charge-doping levels [9,61,62]. FeSe crystals used to be grown by chemical-vapor-transport [86,87], flux-free floating-zone [88], and flux solution methods. These methods are usually hard to tune the chemical stoichiometry.

## 3. Electronic and superconducting properties studied in the hydrothermal single crystals and films

Now we briefly review our recent year's studies of the series of FeSe-based single crystals and films grown by the hydrothermal methods.

### 3.1 Strong electronic two-dimensionality in high-$T_c$ (Li,Fe)OHFeSe single crystal

Fig.3a shows the temperature dependence of the in-plane resistivity, $\rho_{ab}$, for the high-$T_c$ (42 K) $(Li_{0.84}Fe_{0.16})OHFe_{0.98}Se$ single crystal [58], which displays a metallic behavior over the whole measuring temperature range in the normal state. As a measure of the charge transport anisotropy, the ratio of the out-of-plane to in-plane resistivity, $\rho_c/\rho_{ab}$, was found to increase with lowering temperature and reach a high value of 2500 at 50 K. It is obvious that the normal-state electronic property turns out to be highly two dimensional just above $T_c$. Shown in Fig. 3b is the temperature dependence of static magnetic susceptibility, which is slightly dependent on the magnitude of the applied field. In the higher temperature range, all the data can be fitted to a modified Curie-Weiss law $\chi_m = \chi_0 + \chi_{CW}$ (the solid lines), where $\chi_0$ is the Pauli paramagnetic contribution from itinerant charge carriers. A deviation from the Curie-Weiss law is clearly visible below a characteristic $T^*$ (~120 K) of a Hall-coefficient dip feature, coinciding with the upturn in Hall coefficient and change in the resistivity behavior. From the Hall-dip $T^*$ down to the superconducting $T_c$, both the extracted iron-plane magnetic susceptibility (with the Curie-Weiss term subtracted; inset of Fig. 3b) and the in-plane resistivity (inset of Fig. 3a) exhibit a linear temperature dependence, suggesting the presence of two-dimensional antiferromagnetic spin fluctuations in the iron planes.

### 3.2 Phase diagram and electronic phase separation of (Li,Fe)OHFeSe system

The first phase diagram of (Li, Fe)OHFeSe system [63] was based on the powder samples. In a subsequent work [60], we established a more complete phase diagram for the system (Fig. 4), based on a series of the hydrothermal single crystals in the superconducting (SC) and non-superconducting regimes. In some of the SC samples ($T_c < \sim 38$ K, cell parameter $c < \sim 9.27$ Å), we observed a strong drop in the magnetization at an almost constant temperature scale $T_{afm} \sim 125$ K (Fig. 5a, b), indicating the occurrence of antiferromagnetism well above $T_c$. Our analysis of electron energy-loss spectroscopy combined with selected-area electron diffraction confirmed the absence of magnetic impurity phases such as $Fe_3O_4$. Therefore, the antiferromagnetic signal is intrinsic to (Li, Fe)OHFeSe system. Moreover, a positive correlation between the sizes of the antiferromagnetic signal and the Meissner signal was observed (Fig. 5c and d). These experimental results demonstrate the coexistence of an antiferromagnetic state with the superconducting state in (Li, Fe)OHFeSe at $T_c <$ ~38 K and $c < \sim 9.27$ Å. Such coexistence can be explained by electronic phase separation [60], similar to the cases of high-$T_c$ cuprates and iron arsenides. Therefore the electronic phase diagram shown in Fig. 4 provides more information about the electronic states in (Li, Fe)OHFeSe system.

### 3.3 The link between the superconducting and normal-state properties in $Fe_{1-x}Se$ single crystals

The in-plane angular-dependent magnetoresistance (AMR) in the normal state was measured for the hydrothermal $Fe_{1-x}Se$ single crystals [64]. Fig. 6 shows the AMR at a 9 Tesla field for a representative sample with $T_c = 7.6$ K. The AMR displays a two-fold rotational symmetry emerging below a characteristic temperature $T_{sn} \sim 55$K. This anisotropy in AMR is enhanced with decreasing temperature (left panel of Fig. 6). Such an enhancement in charge scatterings was also observed in the temperature-dependent magnetoresistance by an earlier study [89]. Moreover, a downward curvature below $T_{sn} \sim 55$K was observed in our sample in the static magnetization under an in-plane magnetic field of 0.1 T (Fig. 7a) [61]. Such a feature is strongly dependent on the magnitude and direction of the applied field (Fig. 7a vs. b). This suggests that the strong quantum spin frustrations predominate in the iron planes. Although the orbital-nematic order associated with the structural transition at $T_s \sim 90$ K is also of a two-fold rotational symmetry, the obvious downward feature of in-plane static magnetization below $T_{sn} \sim 55$ K, which is far below $T_s$, suggests that the fourfold-rotational-symmetry breaking identified by our AMR measurements is closely related to the frustrated spins with anisotropic magnetic fluctuations. Therefore, a field-induced nematic state of a spin origin emerges below $T_{sn}$.

By summarizing all the data of our samples, we found a remarkable linear relationship between $T_c$ and $T_{sn}$, as shown in Fig. 8. Moreover, the related data of $T_c$ vs. $T_{sn}$ available from literature [89-91] also well satisfy this linear relationship. Namely, the linear relationship between superconducting $T_c$ and characteristic $T_{sn}$ of the field-induced spin-nematic state was observed to cover a wide range from far below to beyond $T_s$. This further suggests that the superconductivity is more likely related to the anisotropic magnetic fluctuations. These results of prototypal FeSe system are consistent with those of intercalated high-$T_c$ (Li,Fe)OHFeSe presented above. It needs to be emphasized that, for nearly stoichiometric FeSe samples with a constant $T_c \sim 9.5$ K, both the spin-nematic ordering and orbital-nematic ordering (associated with the structural transition) happen to coincide with each other at ~ 90 K, as shown in Fig. 8. So it is difficult to distinguish these different ordering states in such samples. Our samples with different $T_c$'s enable the disentanglement of the different states.

Most recently, we have studied the doping dependences of electronic correlation effect [9] and upper critical field behavior [62] in a series of hydrothermal $Fe_{1-x}Se$ single crystals. Particularly in these binary $Fe_{1-x}Se$ samples, the charge-doping level can be tuned simply by the non-stoichiometric $x$ of Fe, from a strong electron dominance at $x \sim 0$ to a strong hole dominance at higher $x$ values. Importantly, we find that superconductivity of FeSe system emerges from the strongly correlated, hole-dominated $Fe_{1-x}Se$ as the non-stoichiometry is reduced to $x \sim 5.3$ % [9]. Interestingly, such an $x$ threshold for superconductivity of the prototypal FeSe is similar to that ($x \sim 5$ % [65]) for high-$T_c$ superconductivity of the intercalated (Li, Fe)OHFeSe sharing the common superconducting FeSe-blocks.

### 3.4 High superconducting critical parameters of un-doped and Mn-doped (Li,Fe)OHFeSe crystals and films

Fig. 9 shows the x-ray diffraction characterization of a representative (Li,Fe)OHFeSe film sample hydrothermally grown on LaAlO$_3$ substrate [16]. The observation of only (00$l$) reflections indicates a single preferred (001) orientation (Fig. 9a). Shown in Fig. 9b is the double-crystal x-ray rocking curve for the (006) Bragg reflection, with a small FWHM = 0.22˚. To our knowledge, this is the best FWHM value observed so far among various iron-based superconductor crystals and films, indicating a high sample quality. The $\phi$-scan of (101) plane shown in Fig. 9c exhibits four successive peaks with an equal interval of 90˚, consistent with the $C_4$ symmetry of the (Li,Fe)OHFeSe film. These results clearly demonstrate an excellent in-plane orientation and epitaxial growth.

High-quality superconducting films can play an important role in the application. Besides the high sample quality, the (Li,Fe)OHFeSe films also display excellent superconducting properties. The temperature dependence of in-plane resistivity is shown in Fig. 10a, with a superconducting zero-resistivity temperature up to 42.4 K. Fig. 10b is the temperature dependences of upper critical field $H_{c2}$ derived from systematic measurements of the in-plane and out-of-plane magnetoresistance. Based on WHH (Werthamer-Helfand-Hohenberg) model, the values of $H_{c2}(0)$ are estimated as 79.5 T and 443 T at magnetic fields perpendicular and parallel to the $ab$ plane, respectively. Moreover, a large critical current density $J_c > 0.5$ MA/cm$^2$ was achieved at ∼20 K, as shown in Fig. 10c. The high superconducting critical parameters are important for practical application. Additionally, as seen from Fig. 11, the critical temperature $T_c$ of (Li$_{0.84}$Fe$_{0.16}$)OHFe$_{0.98}$Se single crystal can be further raised up to a value > 50 K under a pressure of 12.5 GPa, in the superconducting phase II (SC-II) region. The SC-II phase develops with pressure at a critical $P_c = 5$ GPa as the superconducting phase I (SC-I) is gradually suppressed.

Very recently, we have successfully doped Mn into (Li,Fe)OHFeSe films [17]. As seen from Fig. 12a, the $J_c$ value of high-$T_c$ (Li,Fe)OHFeSe film is strongly enhanced by one order of magnitude, from the undoped 0.03 to Mn-doped 0.32 MA/cm$^2$ under 33 T at 5 K. The vortex pinning force density $F_p$ monotonically increases with field up to 106 GN/m$^3$, shown in Fig. 12b. To the best of our knowledge, these values are the records so far among all the iron-based superconductors. Such a superconducting (Li,Fe)OHFeSe film is not only important for the fundamental research, but also promising for high-field application.

### 4. Conclusion

High-quality single crystals and single-crystalline films of iron-based superconductors play an important role in both the basic research and potential application. However, for the FeSe-based superconductor systems reviewed here, by the conventional high-temperature growth it is either hard to obtain the single crystals and films, or not easy to tune the electronic properties. These problems can be overcome by our recently developed soft-chemical hydrothermal growth methods, which are capable of producing the single crystals and films, and tuning the chemical stoichiometry thus the electronic properties. In addition, these methods may be applicable in other layered materials, providing a new route for the exploration of functional materials.

The successful crystal and film growth has enabled systematic studies of the FeSe-based superconductor systems. We have observed a strong electronic two-dimensionality towards $T_c$, and a nearly linear extracted magnetic susceptibility as well as a linear in-plane resistivity, both emerging from a characteristic $T^*$ (~120 K) of a Hall-coefficient dip feature down to $T_c$, in high-$T_c$ intercalated (Li,Fe)OHFeSe system. We have also observed a linear relationship between $T_c$ and a characteristic $T_{sn}$ of a field-induced spin nematicity in prototypal FeSe system. These results suggest the presence of magnetic fluctuations in the iron planes and their relevance to superconductivity. Importantly, we have found that superconductivity of the prototypal FeSe emerges from the strongly correlated, hole-dominated $Fe_{1-x}Se$ at a non-stoichiometric $x$ similar to that for the high-$T_c$ superconductivity of the FeSe-intercalate (Li, Fe)OHFeSe. An electronic phase diagram has been established for (Li, Fe)OHFeSe system, with the coexistence of antiferromagnetism and superconductivity explained by electronic phase separation. On the other hand, the high superconducting critical current density achieved in Mn-doped high-$T_c$ (Li,Fe)OHFeSe film is promising for high-field application. These FeSe-based superconductor systems deserve further experimental and theoretical studies, in both aspects of the underlying physics and potential application.

**Conflict of Interest**

*The authors declare that the research was conducted in the absence of any commercial or financial relationships that could be construed as a potential conflict of interest.*

**Author contributions**

All authors contribute to the writing of the manuscript.

**Funding**

This work was supported by National Natural Science Foundation of China (Nos. 11834016 and 11888101), the National Key Research and Development Program of China (Grant Nos. 2017YFA0303003, 2016YFA0300300), and the Strategic Priority Research Program and Key Research Program of Frontier Sciences of the Chinese Academy of Sciences (Grant Nos. XDB25000000, QYZDY-SSW-SLH001)

**ACKNOWLEDGMENTS**

We are very thankful to all our collaborators for their valuable scientific contributions in the past six years, especially Huaxue Zhou, Dongna Yuan, Yulong Huang, Yiyuan Mao, Shunli Ni, Jinpeng Tian, Dong Li, and Peipei Shen for sample preparation and characterization. We are also very grateful to Kui Jin, Jie Yuan, Wei Hu, and Zhongpei Feng for electrical transport measurements and insightful discussions; Jinguang Cheng and Jianping Sun for high-pressure research; Zian Li, Huaixin Yang, and Jianqi Li for TEM & EELS studies; Guangming Zhang and Zhenyu Zhang for theoretical support; Xingjiang Zhou and Lin Zhao for ARPES studies; Donglai Feng and Tong Zhang for STM studies; Li Pi, Chuanying Xi, Zhaosheng Wang, and J. Wosnitza for high-field studies; Rustem Khasanov, Zurab Guguchia, and Alex Amato for µSR studies. We also thank Ping Zheng, Shaokui Su, and Lihong Yang for technical supports.

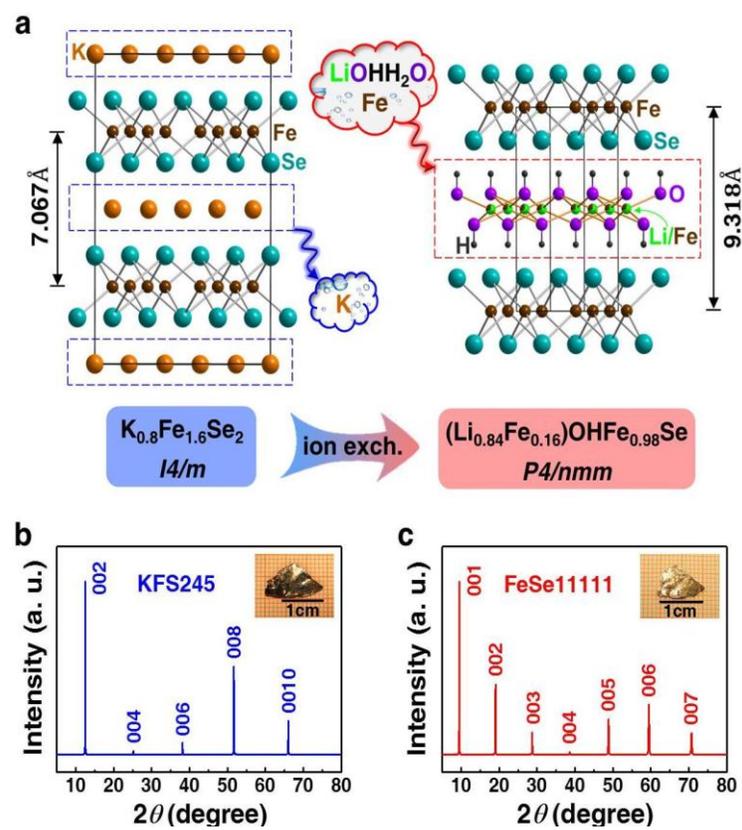

Figure 1 Illustration of hydrothermal ion-exchange growth of (Li,Fe)OHFeSe crystals [58].

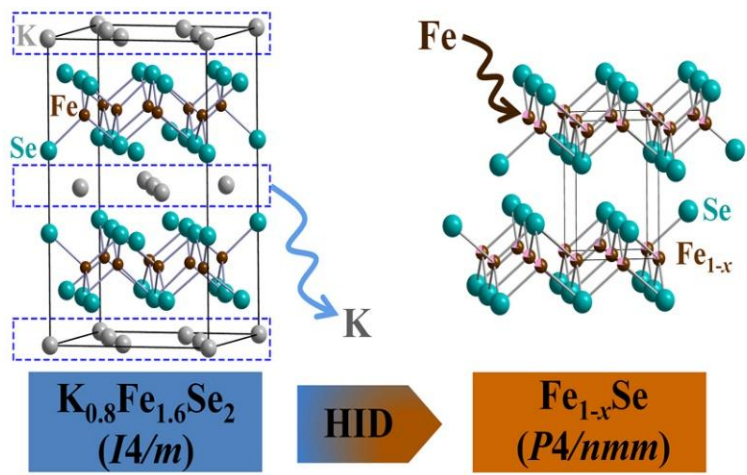

Figure 2 Scematic llutration of the hydrothermal ion-deintercalation method. During the HID process, $Fe_{1-x}Se$ single crystals are derived from the readily obtainable phase-pure matrix single crystals of $K_{0.8}Fe_{1.6}Se_2$. The original interlayer K ions and Fe vacancies (20 % in amount) in $K_{0.8}Fe_{1.6}Se_2$ were completely de-intercalated and substantially reduced, respectively, yielding the target single crystals of phase-pure $Fe_{1-x}Se$ [9,61,62].

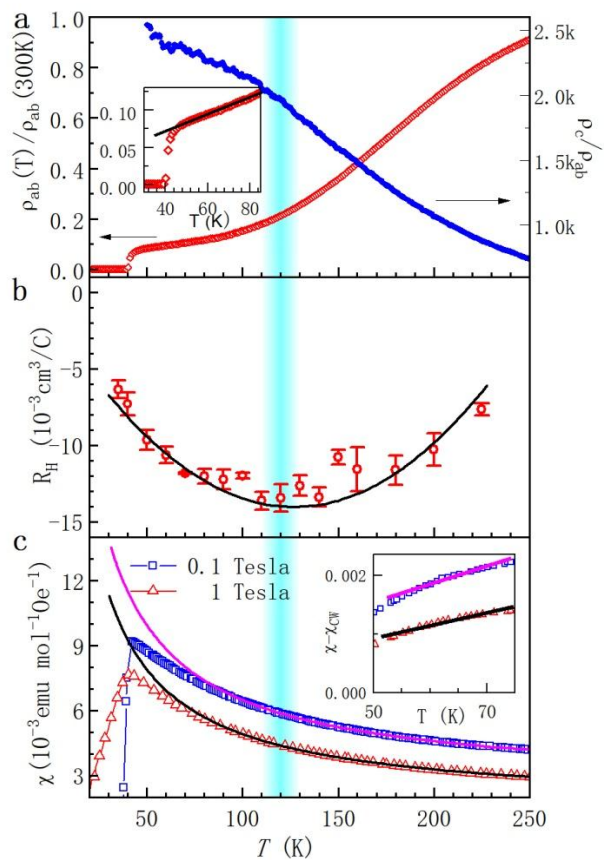

Figure 3 The electrical transport and magnetic properties of $(Li_{0.84}Fe_{0.16})OHFe_{0.98}Se$ single crystal [58]. (a) The in-plane electrical resistivity and the ratio of out-of-plane to in-plane resistivity as functions of temperature. The inset shows the linear resistivity below the Hall-dip $T^*$ down to $T_c$. (b) The temperature dependence of in-plane Hall coefficient shows a dip feature around $T^* \sim 120$ K. (c) The temperature dependencies of static magnetic susceptibility under magnetic fields along $c$-axis. A deviation from the Curie-Weiss law is clearly visible below the Hall-dip $T^*$. After subtracting the Curie-Weiss term (the solid fitted curves) from the $(Li_{0.84}Fe_{0.16})OH$-blocks, a nearly linear magnetic susceptibility from the FeSe-blocks is obvious (the inset).

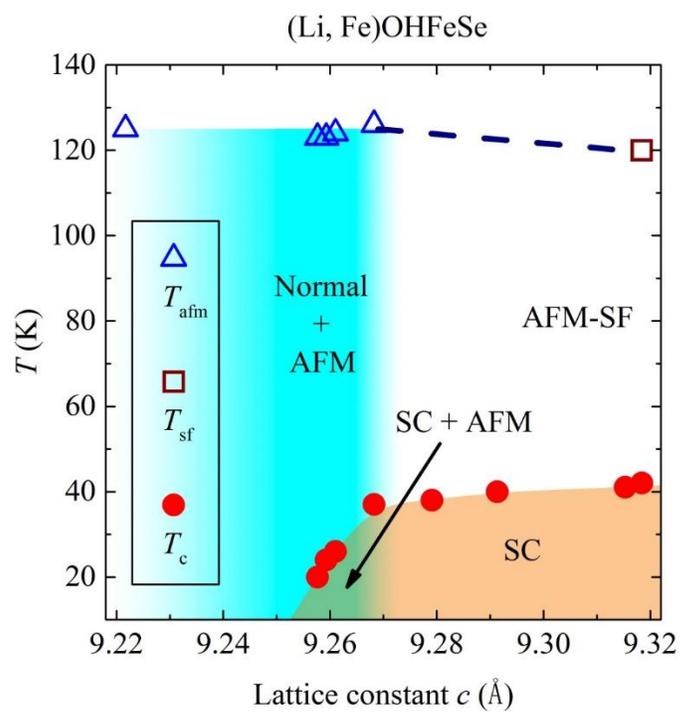

Figure 4 Electronic phase diagram of (Li,Fe)OHFeSe system [60,63].

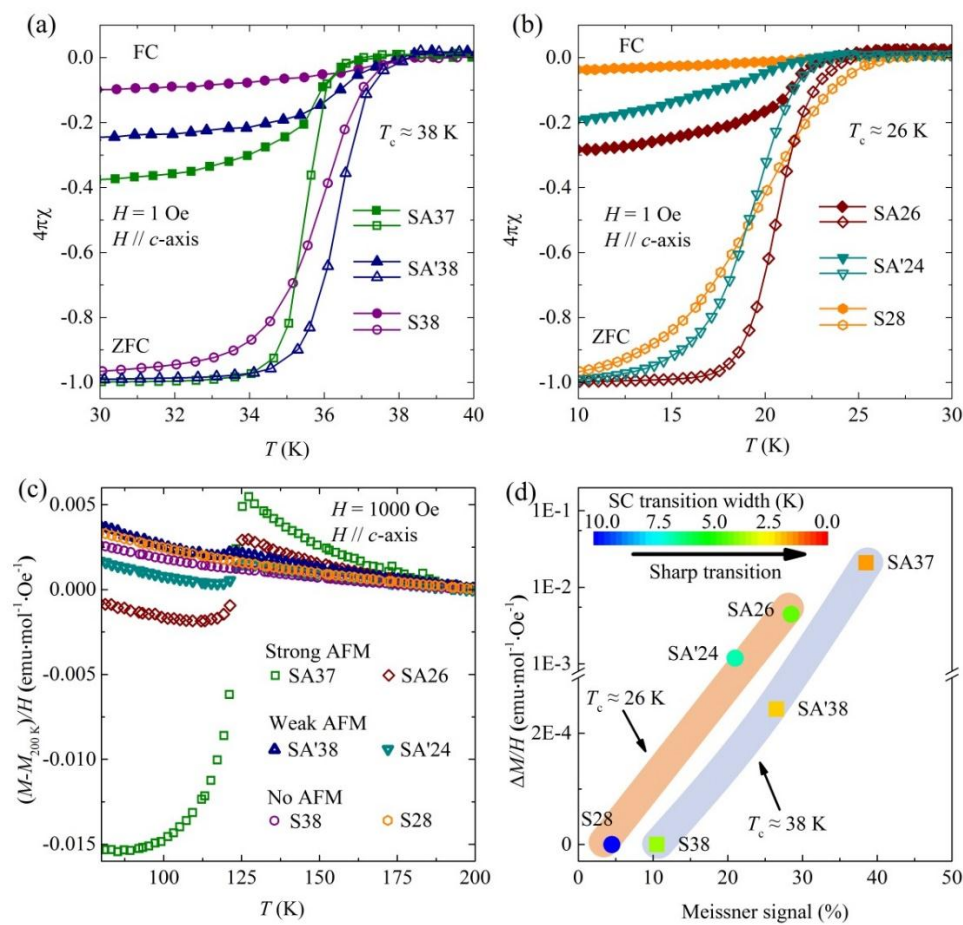

Figure 5 (a) and (b) Temperature dependence of static magnetic susceptibility near the superconducting transitions, for the two sets of superconducting (Li,Fe)OHFeSe single crystals. (c) Antiferromagnetic (AFM) transition at ~125 K is detectable for the superconducting (Tc < ~38K) samples and non-superconducting samples. (d) The corresponding AFM signal size and the SC Meissner signal size are positively correlated [60].

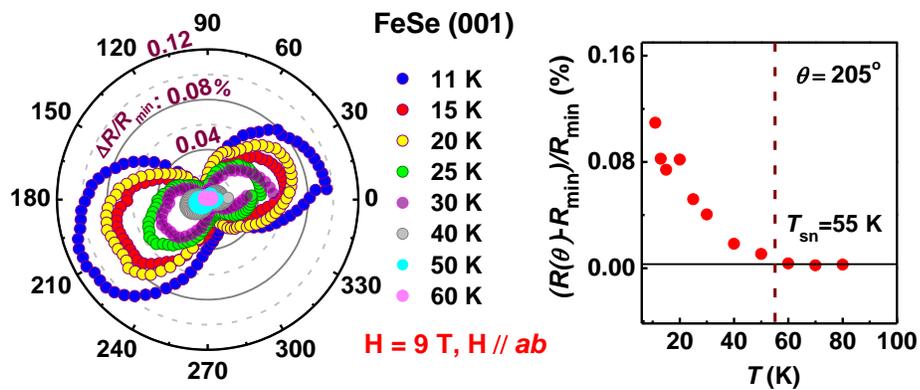

Figure 6 Temperature dependences of the angular-dependent magnetoresistance of FeSe crystal ($T_c$ = 7.6 K), showing the twofold rotational symmetry below $T_{sn}$ ~55K [64].

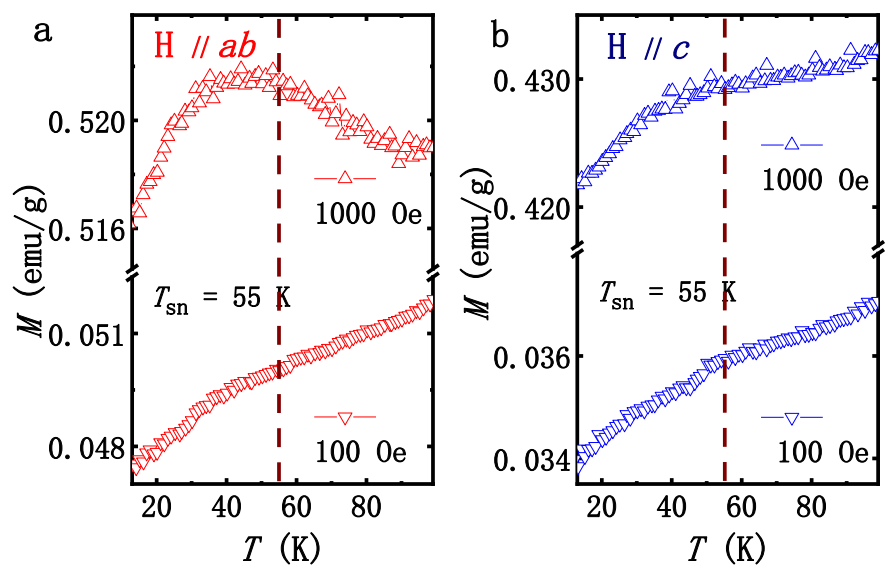

Figure 7 Temperature dependence of the static magnetization around 55 K under the in-plane and out-of-plane fields for the FeSe crystal shown in Fig. 6. [61]

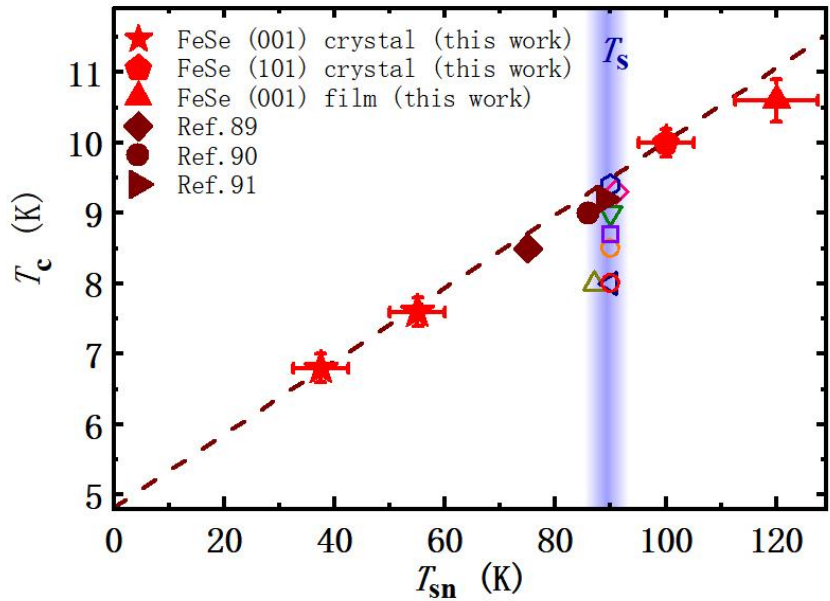

Figure 8 The universal linear relationship between the superconducting transition temperature ($T_c$) and the field-induced spin-nematic ordering temperature ($T_{sn}$) among various FeSe samples FeSe samples (the solid symbols) [64]. The hollow symbols in the vertical blue-shaded area represent the structure phase transition temperatures by the x-ray or neutron diffractions on various FeSe samples of different Tc's [30,52,55,86,92-94].

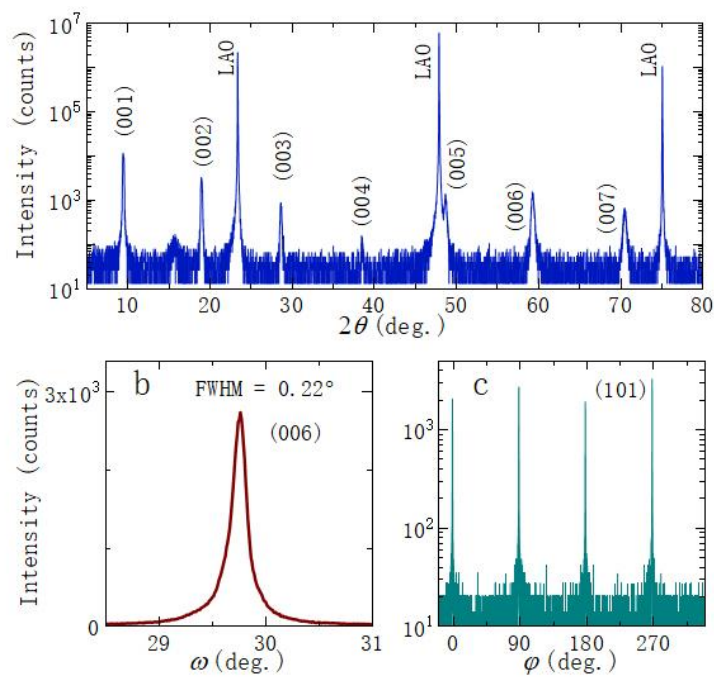

Figure 9 XRD characterizations of the (Li,Fe)OHFeSe film on the LaAlO$_3$ (LAO) substrate. (a) The $\theta$–$2\theta$ scan shows only (00$l$) peaks. (b) The rocking curve of (006) reflection with an FWHM of 0.22˚. (c) The $\phi$-scan of the (101) plane. The uniform 4-fold symmetry reveals an excellent epitaxial growth [16].

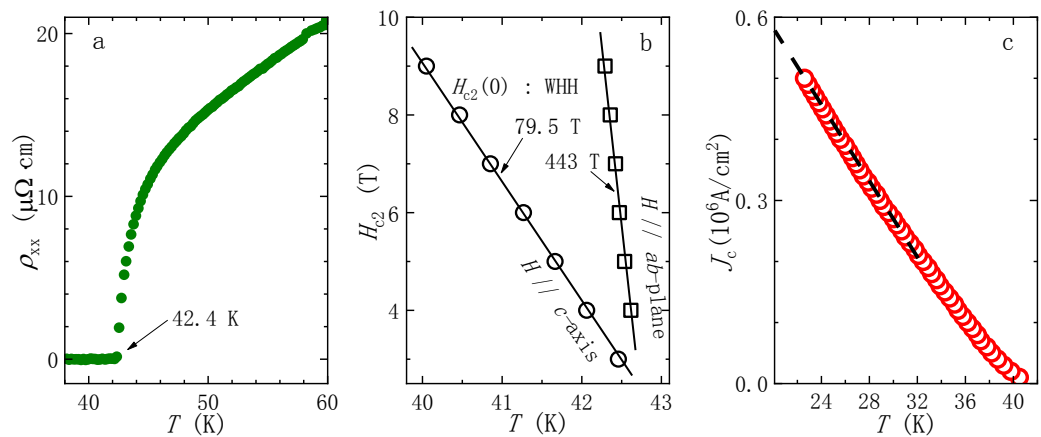

Figure 10 High superconducting critical parameters for (Li,Fe)OHFeSe film. (a) Temperature dependence of in-plane resistivity, with the onset of zero resistivity at 42.4 K. (b) Temperature dependence of $H_{c2}(T)$ along the $c$-axis (circle) and within the $ab$ plane (square). (c) The temperature dependence of $J_c$, exceeding 0.5MA/cm$^2$ at 20K [16].

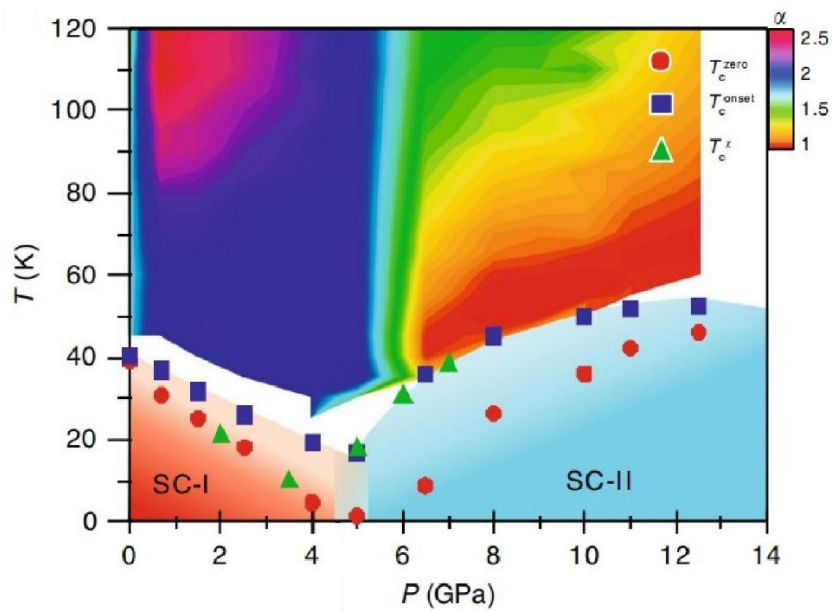

Figure 11 Temperature-pressure phase diagram of $(Li_{0.84}Fe_{0.16})OHFe_{0.98}Se$ single crystal [70]. Pressure-dependence of $T_c$ and a contour color plot of the normal-state resistivity exponent $\alpha$ up to 12.5 GPa are shown.

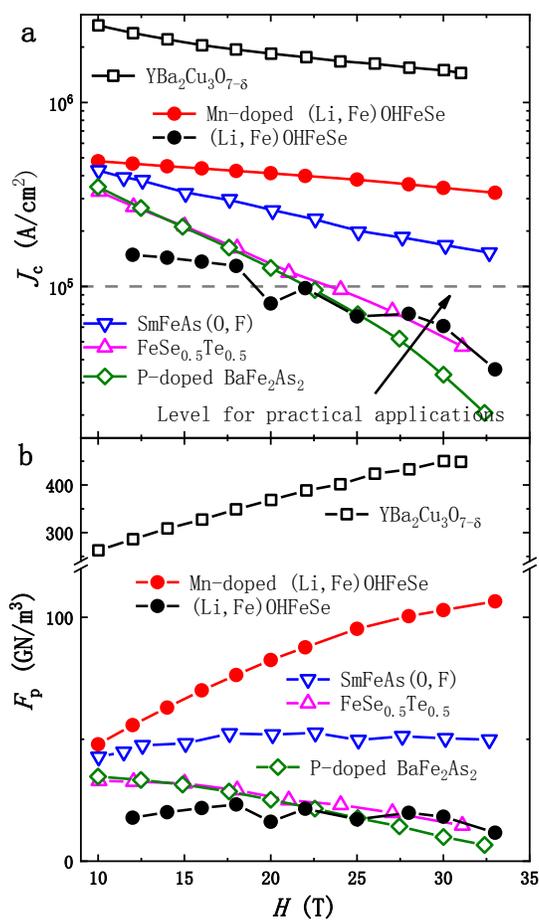

Figure 12 Magnetic field dependence of $J_c$ (a) and $F_p$ (b) of several superconductors [17], including Mn-doped and pure (Li,Fe)OHFeSe films at 5 K, SmFeAs(O,F) films [95], FeSe$_{0.5}$Te$_{0.5}$ films [96], P-doped BaFe$_2$As$_2$ films [97], and YBa$_2$Cu$_3$O$_{7-\delta}$ wires [98] at 4.2 K under c-axis fields.